\documentclass[a4paper,11pt]{article}
\usepackage{pos}

\title{CP Violation and Mixing in Beauty Sector: \\A Theoretical Overview}
\author*[a]{Eleftheria Malami}
\affiliation[a]{Nikhef,\\
Science Park 105, NL-1098 XG Amsterdam, Netherlands}

\emailAdd{emalami@nikhef.nl}
\abstract{We provide an overview of the latest progress in the study of CP violation and mixing in the $B$ meson decays. Studying the $B^0_d \to J/\psi K^0_S$ and $B^0_s \to J/\psi \phi$ decays, we focus on the determination of the mixing phases $\phi_d$ and $\phi_s$, including hadronic uncertainties with the help of the data. CP violation in the non-leptonic decays $B \to \pi K$ and  $B^0_s \to D_s^{\mp} K^{\pm}$ indicate puzzling patterns. Could these puzzles indicate New Physics?}

\FullConference{%
  The Tenth Annual Conference on Large Hadron Collider Physics - LHCP2022\\
  16-20 May 2022\\
  online
}

\begin{document}
\maketitle
\section{Introduction}

CP Violation and Mixing in the beauty sector are important for testing the Standard Model (SM) and a good probe for physics beyond the SM. The neutral $B$ mixing efffects in the SM are described by loop diagrams, called boxes. The CP-violating phases  $\phi_d$ and $\phi_s$, which are associated  with $B^0_q-\bar{B}^0_q$, mixing $(q=d,s)$ are important for searching for hints of New Physics (NP). The SM predictions of the mixing phases strongly depend on the input parameters and especially the apex of the Unitarity Triangle (UT). As tensions arise between measurements of the elements of the Cabibbo–Kobayashi–Maskawa (CKM) matrix \cite{Cabibbo:1963yz,Kobayashi:1973fv}, leading to different pictures for the allowed NP parameter space in the $B_q$-meson mixing \cite{DeBruyn:2022zhw}, it becomes essential to investigate their origin and eventually resolve them. Here, we focus on $B$ meson decays to determine the CP-violating phases and explore phenomena that comes from mixing-induced CP violation, which involve interference effects caused by $B^0_q-\bar{B}^0_q$ mixing. Interestingly, we also encounter puzzling patterns which follow from current data.

\section{Mixing in $B^0_d \to J/\psi K^0_S$ and $B^0_s \to J/\psi \phi$ decays}
The $B^0_d \to J/\psi K^0_S$ and $B^0_s \to J/\psi \phi$ modes, which are considered ``golden modes'', are characterised by tree topologies as well as penguin topologies, which are doubly Cabibbo suppressed. These penguin contributions are related to non-perturbative effects, therefore it is difficult to calculate them reliably. In order to handle them, one has to use control channels as well as experimental data. The key feature in these control channels is that the penguin effects are not anymore doubly Cabibbo suppressed.

The decay amplitude of the $B^0_d \to J/\psi K^0_s$ channel can be written in the form \cite{Fleischer:1999nz}
\begin{equation}\label{eq:Amp_BdJpsiKS}
    A(B_d^0\to J/\psi K^0) = \left(1-\frac{1}{2}\lambda^2\right)\mathcal{A}'\left[1+\epsilon a'e^{i\theta'}e^{i\gamma}\right]\:,\qquad
    \epsilon \equiv \frac{\lambda^2}{1-\lambda^2} \approx 0.052\:,
\end{equation}
where $\lambda=|V_{us}|$, $\mathcal{A}'$ is the hadronic amplitude and the term $a'e^{i\theta'}$ denotes the penguin parameters. The effective mixing phase is determined as \cite{Faller:2008zc,Faller:2008gt}:
\begin{equation}
  \sin\left(\phi_q^{\text{eff}}\right)
     = \sin\left(\phi_q^{\text{SM}} + \phi_q^{\text{NP}} + \Delta\phi_q\right) = {\eta_f \mathcal{A}_{\text{CP}}^{\text{mix}}(B_q\to f)} / {\sqrt{1 - \left(\mathcal{A}_{\text{CP}}^{\text{dir}}(B_q\to f)\right)^2}} \:,
\end{equation}
where $\eta_f$ is the CP eigenvalue of the final state $f$ and the $\Delta\phi_q$ is a hadronic phase shift associated with penguin topologies.

A state-of-the-art analysis has been presented in \cite{Barel:2020jvf} and the SU(3) flavour symmetry of strong interaction has been applied, which allows us to relate the parameters of $\bar{b} \to \bar{s} c \bar{c}$ to those of $\bar{b} \to \bar{d} c \bar{c}$ transitions (thus the hadronic parameters of the control channels):
\begin{equation}\label{eq:su3_relation}
a' e^{i \theta'}= a e^{i \theta}\:.
\end{equation}

Utilising Eq.~\ref{eq:su3_relation} the $B^0_d \to J/\psi K^0_S$ decay is related to its partner control channels $B^0_s \to J/\psi K^0_S$ and  $B^0_d \to J/\psi \pi^0$ and similarly, the $B^0_s \to J/\psi \phi$ decay to $B^0_d \to J/\psi \rho^0$ channel \cite{Faller:2008zc,DeBruyn:2010hh,DeBruyn:2014oga,Barel:2020jvf}. The mixing-induced CP asymmetries of all these five $B^0_q \to J/\psi X$ channels depend on the $\phi_q$ phases. In order to properly take into account the dependencies between $\phi_d$, $\Delta \phi_d$, $\phi_s$ and $\Delta \phi_s$, a simultaneous analysis between these five decays has been proposed \cite{Barel:2020jvf}. This allows the extraction of the relevant hadronic parameters and the mixing phases \cite{Barel:2022wfr}:
\begin{equation}\label{eq:results_phiq}
    \phi_d = \left(44.4_{-1.5}^{+1.6}\right)^{\circ}  \:, \qquad
      \phi_s = -0.074_{-0.024}^{+0.025} = \left(-4.2 \pm 1.4\right)^{\circ}\:,
\end{equation}
taking the penguin effects directly into account. Future scenarios show that New Physics (NP) contributions to $\phi_s$ could be established with more than $5\sigma$ in the ultra high-precision era. Discovering NP in the $B_s$ system is possible but only if there are improvements in the measurements of the CP asymmetries of all the five $B^0_q \to J/\psi X$ channels. Regarding $\phi_d$, revealing NP effects is limited by the knowledge of the UT apex, determined through the angle $\gamma$ and the side $R_b$, which strongly depends on the choice of the CKM matrix elements $|V_{ub}|$ and $|V_{cb}|$. In the future, it is necessary to improve the precision on $R_b$ and resolve the tension between the inclusive and exclusive determination of these CKM elements from semileptonic $B$ decays \cite{Barel:2020jvf,Barel:2022wfr,DeBruyn:2022zhw}.

Another important topic is the determination of the effective colour-suppression factor $a_2$  in the cleanest possible way from the data. For this purpose, a new strategy has been proposed using semileptonic decays. Hence, for the $B^0_d \to J/\psi \pi^0$ decay (and $B^0_s \to J/\psi K^0_S$) one may use the semileptonic channel $B^0_d \to \pi^- \ell^+ \nu_{\ell}$ (and $B^0_s \to K^- \ell^+ \nu_{\ell}$, respectively) to create the ratio \cite{Barel:2020jvf}:
\begin{equation}\label{eq:SL_ratio}
    R_d^\pi %\equiv \frac{\Gamma(B^0_d \to J/\psi \pi^0)}{d\Gamma/dq^2|_{q^2=m_{J/\psi}^2}(B^0_d \to \pi^- \ell^+ \nu_{\ell})}
    = {\mathcal{B}(B^0_d \to J/\psi \pi^0)}/ {\left[d\mathcal{B}/dq^2|_{q^2=m_{J/\psi}^2}(B^0_d \to \pi^- \ell^+ \nu_{\ell}) \right]}\:,
\end{equation}
where the CP-averaged branching fraction of the $B^0_d \to J/\psi \pi^0$ decay is given by:
\begin{align}
   2 \: \mathcal{B}(B_d^0\to J/\psi\pi^0) =  & 
   \: \tau_{B_d} \: \frac{G_{\mathrm F}^2}{32 \pi} |V_{cd}V_{cb}|^2 \: m_{B_d}^3
   \left[ f_{J/\psi}  f_{B_d\to \pi}^+(m_{J/\psi}^2) \right]^2
    \left[\Phi\left(\frac{m_{J/\psi}}{m_{B_d}},\frac{m_{\pi^0}}{m_{B_d}}\right)\right]^3 \nonumber \\
    & \times (1 - 2 a\cos\theta\cos\gamma + a^2) \times \left[ a_2 (B_d^0\to J/\psi\pi^0) \right]^2.
\end{align}
The hadronic form factors drop out in Eq.~\ref{eq:SL_ratio}, yielding $a_2=0.363^{+0.066}_{-0.079}$, thus allowing a theoretically clean extraction of the $a_2$ value. This result agrees with expectation from naive factorisation, $a_2=0.21 \pm 0.05$ \cite{Buras:1998us}. This is a very interesting finding suggesting that factorisation is working better than it was expected in this category of decays.

\section{CP Violation in $B \to \pi K$ and $B^0_s \to D_s^{\mp} K^{\pm}$ Decays}
Over the years, the $B \to \pi K$ puzzle has received a lot of attention and remains an intriguing problem \cite{Fleischer:1996bv, Buras:2003dj, Gronau:2008gu, Fleischer:2008wb, Fleischer:2018bld}. These decays are dominated by gluonic (QCD) loop diagrams (penguins) but electro-weak penguins (EWP) play also an important role. One of the most important channels is the $B^0_d \to \pi^0 K_S$, which is the only mode which exhibits mixing-induced CP violation. 

A state-of-the-art analysis of the correlation between the mixing-induced and the direct CP asymmetries has shown that the tension with the data has become even stronger compared to the previous results. In order to resolve this puzzle either the $B^0_d \to \pi^0 K_S$ data should change or NP might be present. In the first case, reducing the branching ratio by $2.5 \sigma$ would lead to a consistent picture with the SM. On the other hand, if NP effects are present, then a modified EWP sector is a possible interesting candidate. A new strategy has been proposed in order to determine the EWP parameters, using both neutral and charged $B \to \pi K$ decays \cite{Fleischer:2018bld}. In the future, it will be interesting to measure CP violation with much higher precision, in particular the $B^0_d \to \pi^0 K_S$ mixing-induced CP asymmetry and hopefully to clarify the situation.

Another surprising situation arises in the $B^0_s \to D_s^{\mp} K^{\pm}$ system \cite{Fleischer:2021cct,Fleischer:2021cwb}. These are pure tree decays and theoretically clean, making them key players in the testing of the SM. The first puzzling pattern is related to the value of the angle $\gamma$ of the UT \cite{LHCb:2017hkl}. Through interference effects with the $B^0_s-\bar{B}^0_s$ mixing, the CP asymmetry parameters allow the determination of the term $\phi_s + \gamma$, thus $\gamma$. The result is in tension with the SM at the $3 \sigma$ level. Taking into account that the experimental $\phi_s$ value includes NP effects in $B^0_s - \bar{B}^0_s$ mixing, possible new contributions should enter at the decay amplitude level. 

Such NP effects should also appear in the corresponding branching ratios. Extracting individual branching ratios and comparing with the SM, a second intriguing puzzle arises. The case gets even more exciting as similar patterns are also observed in the branching ratios of other $B$ meson decays which have similar dynamics. Utilising information from semileptonic decays, the phenomenological colour factors $|a_1|$, characterising colour-allowed tree decays, can be extracted in a way that minimises the dependence on CKM matrix elements and hadronic form factors. The tensions between the theoretical predictions and the experimental values are up to $4.8\sigma$ level. 

In view of these intriguing puzzles, a new model independent description has been suggested in order to include NP effects. Introducing the NP parameters $\bar{\rho}, \rho$, which denote the strength of NP contributions with respect to the SM amplitudes, as well as $\bar{\delta}, \delta$ and $\bar{\varphi}, \varphi$, which represent the strong and the weak CP-violating phases, respectively, the amplitudes are parametrised. The expression of the observables $\xi$ and $\bar{\xi}$ (which measure the strength of interference effects) has been generalised as well as the assumption $C +\bar{C} = 0$, which was used by LHCb Collaboration. Therefore, the general expression takes the form:
\begin{equation}
\xi \times \bar{\xi}  = \sqrt{1-2\left[\frac{C+\bar{C}}{\left(1+C\right)\left(1+\bar{C}\right)}
\right]}e^{-i\left[2 (\phi_s + \gamma)+\gamma_{\text{eff}}\right]} \qquad {\text{with}} \qquad \gamma_{\rm eff}\equiv \gamma + \gamma_{\text{NP}},
\end{equation}
where $\gamma$ enters with a shift $\gamma_{\text{NP}}=f(\bar{\rho}, \rho, \bar{\varphi}, \varphi)$, which is due to CP-violating NP contributions, resulting in an effective angle $\gamma_{\rm eff}$. Applying the strategy to the current data, correlations between the NP parameters are determined. Interestingly, it is shown that NP contributions at the $30 \%$ level could accommodate the data.

\section{Conclusions}
The decays $B^0_d \to J/\psi K^0_S$ and $B^0_s \to J/\psi \phi$ are the central modes for the determination of the CP violating phases $\phi_d$ and $\phi_s$, which are associated with the neutral $B$ mixing phenomena. We are entering a phase where hadronic uncertainties have to be included to match the experimental precision. A strategy has been developed of including these corrections with the help of data. Other non-leptonic decays indicate puzzling patterns. Regarding the $B \to \pi K$ system, key quantity is the mixing-induced CP asymmetry of the $B^0_d \to \pi^0 K_S$ channel. Another interesting system, which has different dynamics though, is given by $B^0_s \to D_s^{\mp} K^{\pm}$ decays, showing puzzles both in CP violation and in the branching ratios. Consistent patterns arise in other $B_{(s)}$ modes with similar dynamics, making the situation even more interesting. Could these puzzles in the non-leptonic decays indicate NP? It will be exciting to see how the data will evolve in the future high precision era.

\section*{Acknowledgements}
I would like to thank R. Fleischer, K. De Bruyn, K. K. Vos, P. van Vliet, R. Jaarsma and M. Z. Barel for interesting collaborations and useful discussion. I would also like to thank the organisers of the "LHCP2022 Conference" for the invitation.

\end{document}